\documentclass[11pt]{article}
\usepackage{latexsym}
\input epsf.tex

\def\AFOUR{%
\setlength{\textheight}{8.5in}%
\setlength{\textwidth}{5.75in}%
\setlength{\topmargin}{-0.375in}%
\hoffset=-.5in%
\renewcommand{\baselinestretch}{1.17}%
\setlength{\parskip}{6pt plus 2pt}%
}

\AFOUR                                           

\expandafter\ifx\csname amssym.def\endcsname\relax \else\endinput\fi
\expandafter\edef\csname amssym.def\endcsname{%
       \catcode`\noexpand\@=\the\catcode`\@\space}
\catcode`\@=11
\def\undefine#1{\let#1\undefined}
\def\newsymbol#1#2#3#4#5{\let\next@\relax
 \ifnum#2=\@ne\let\next@\msafam@\else
 \ifnum#2=\tw@\let\next@\msbfam@\fi\fi
 \mathchardef#1="#3\next@#4#5}
\def\mathhexbox@#1#2#3{\relax
 \ifmmode\mathpalette{}{\m@th\mathchar"#1#2#3}%
 \else\leavevmode\hbox{$\m@th\mathchar"#1#2#3$}\fi}
\def\hexnumber@#1{\ifcase#1 0\or 1\or 2\or 3\or 4\or 5\or 6\or 7\or 8\or
 9\or A\or B\or C\or D\or E\or F\fi}

\font\tenmsa=msam10
\font\sevenmsa=msam7
\font\fivemsa=msam5
\newfam\msafam
\textfont\msafam=\tenmsa
\scriptfont\msafam=\sevenmsa
\scriptscriptfont\msafam=\fivemsa
\edef\msafam@{\hexnumber@\msafam}
\mathchardef\dabar@"0\msafam@39
\def\dashrightarrow{\mathrel{\dabar@\dabar@\mathchar"0\msafam@4B}}
\def\dashleftarrow{\mathrel{\mathchar"0\msafam@4C\dabar@\dabar@}}

\def\ulcorner{\delimiter"4\msafam@70\msafam@70 }
\def\urcorner{\delimiter"5\msafam@71\msafam@71 }
\def\llcorner{\delimiter"4\msafam@78\msafam@78 }
\def\lrcorner{\delimiter"5\msafam@79\msafam@79 }
\def\yen{{\mathhexbox@\msafam@55}}
\def\checkmark{{\mathhexbox@\msafam@58}}
\def\circledR{{\mathhexbox@\msafam@72}}
\def\maltese{{\mathhexbox@\msafam@7A}}
\def\circledS{{\mathhexbox@\msafam@73}}

\font\tenmsb=msbm10
\font\sevenmsb=msbm7
\font\fivemsb=msbm5
\newfam\msbfam
\textfont\msbfam=\tenmsb
\scriptfont\msbfam=\sevenmsb
\scriptscriptfont\msbfam=\fivemsb
\edef\msbfam@{\hexnumber@\msbfam}
\def\Bbb#1{{\fam\msbfam\relax#1}}
\def\widehat#1{\setbox\z@\hbox{$\m@th#1$}%
 \ifdim\wd\z@>\tw@ em\mathaccent"0\msbfam@5B{#1}%
 \else\mathaccent"0362{#1}\fi}
\def\widetilde#1{\setbox\z@\hbox{$\m@th#1$}%
 \ifdim\wd\z@>\tw@ em\mathaccent"0\msbfam@5D{#1}%
 \else\mathaccent"0365{#1}\fi}
\font\teneufm=eufm10
\font\seveneufm=eufm7
\font\fiveeufm=eufm5
\newfam\eufmfam
\textfont\eufmfam=\teneufm
\scriptfont\eufmfam=\seveneufm
\scriptscriptfont\eufmfam=\fiveeufm
\def\frak#1{{\fam\eufmfam\relax#1}}

\csname amssym.def\endcsname

\makeatletter
\def\section{\@startsection {section}{1}{\z@}{-3.5ex plus -1ex minus
 -.2ex}{2.3ex plus .2ex}{\large\sc}}
\def\subsection{\@startsection{subsection}{2}{\z@}{-3.25ex plus -1ex minus
 -.2ex}{1.5ex plus .2ex}{\normalsize\sc}}
\makeatother

\makeatletter
\@addtoreset{equation}{section}

\makeatother

\newcommand{\nc}{\newcommand}
\newcommand{\rnc}{\renewcommand}

\nc{\subs}[1]{{\vspace*{0.5cm}}%
{\noindent\underline{\small\sc #1}}{\addcontentsline{toc}{subsubsection}{#1}}%
{\vspace*{0.3cm}}}

\nc{\subss}[1]{{\vspace*{0.5cm}}%
{\noindent\underline{\small\sc #1}}%
{\vspace*{0.3cm}}}

\nc{\chap}[1]{{\clearpage}%
\begin{center}%
{\noindent\underline{\large\sc #1}}{\addcontentsline{toc}{section}{#1}}%
\end{center}%
{\vspace*{0.3cm}}}

\nc{\be}{\begin{equation}}
\nc{\ee}{\end{equation}}
\nc{\bea}{\begin{eqnarray}}
\nc{\eea}{\end{eqnarray}}

\nc{\trac}[2]{{\textstyle\frac{#1}{#2}}}

\nc{\ex}[1]{\mbox{e}^{\,\textstyle#1}}

\nc{\CC}{\Bbb{C}}
\nc{\HH}{\Bbb{H}}
\nc{\PP}{\Bbb{P}}
\nc{\RR}{\Bbb{R}}
\nc{\ZZ}{\Bbb{Z}}
\nc{\II}{\Bbb{I}}
\nc{\EE}{\Bbb{E}}

\rnc{\a}{\alpha}
\rnc{\b}{\beta}
\rnc{\d}{\delta}
\nc{\ga}{\gamma}
\nc{\la}{\lambda}
\nc{\f}{\phi}
\nc{\p}{\psi}
\nc{\e}{\eta}
\rnc{\c}{\chi}
\nc{\eps}{\epsilon}
\nc{\om}{\omega}
\nc{\Om}{\Omega}

\nc{\symx}{\circledS}
\newsymbol\smallsmile 1360
\newsymbol\smallfrown 1361
\nc{\ad}{\mathop{\mbox{ad}}\nolimits}
\nc{\tr}{\mathop{\mbox{tr}}\nolimits}
\nc{\Tr}{\mathop{\mbox{Tr}}\nolimits}
\nc{\Det}{\mathop{\mbox{Det}}\nolimits}
\rnc{\det}{\mathop{\mbox{det}}\nolimits}
\nc{\rk}{\mathop{\mbox{rk}}\nolimits}
\nc{\del}{\partial}
\nc{\diag}{\mathop{\mbox{diag}}\nolimits}
\nc{\ra}{\rightarrow}
\nc{\Ra}{\Rightarrow}
\nc{\LRa}{\Leftrightarrow}
\nc{\lra}{\leftrightarrow}
\nc{\ot}{\otimes}
\rnc{\ss}{\subset}
\nc{\nul}{\noindent\underline}
\nc{\non}{\nonumber\\}
\nc{\mat}[4]{\left(\begin{array}{cc}#1&#2\\#3&#4\end{array}\right)}
\rnc{\lg}{\frak{g}}
\nc{\G}[3]{\Gamma^{#1}_{\;{#2}{#3}}}
\nc{\nam}{\nabla_{\mu}}
\nc{\nan}{\nabla_{\nu}}
\nc{\dx}{\dot{x}}
\nc{\dxl}{\dot{x}^{\la}}
\nc{\dxm}{\dot{x}^{\mu}}
\nc{\dxn}{\dot{x}^{\nu}}
\nc{\ddx}{\ddot{x}}
\nc{\ddxm}{\ddot{x}^{\mu}}
\nc{\ddxn}{\ddot{x}^{\nu}}
\nc{\dxi}{\dot{\xi}}
\nc{\ddxi}{\ddot{\xi}}

\begin{document}


\rightline{SISSA/53/2003/EP}
\vspace{1in}
\begin{center}
{\Large\sc G\"odel, Penrose, anti--Mach:}\\[.3in]
{\Large\sc Extra~Supersymmetries~of~Time-Dependent~Plane~Waves}
\end{center}
\vspace{0.7cm}
\begin{center}
{\large\sc
Matthias Blau\footnote{e-mail: {\tt mblau@ictp.trieste.it}}, 
Patrick Meessen\footnote{e-mail: {\tt meessen@sissa.it}}, 
Martin O'Loughlin\footnote{e-mail: {\tt loughlin@sissa.it}}\\[.3in]
SISSA\\
Via Beirut 4\\
Trieste, Italy}
\end{center}

\begin{center}Abstract\end{center}

We prove that M-theory plane waves with extra supersymmetries are necessarily
homogeneous (but possibly time-dependent), and we show by explicit
construction that such time-dependent plane waves can admit extra
supersymmetries. To that end we study the Penrose limits of G\"odel-like
metrics, show that the Penrose limit of the M-theory G\"odel metric
(with 20 supercharges) is generically a time-dependent homogeneous plane wave of
the anti-Mach type, and display the four extra Killings spinors in
that case. We conclude with some general remarks on the Killing spinor
equations for homogeneous plane waves.

\newpage
\vspace{0.5cm}
\begin{small}
\tableofcontents
\end{small}

\newpage
\setcounter{footnote}{0}

\section{Introduction}

The results of \cite{bfhp1,rm,bfhp2,bmn} have led to a renewed
interest in various aspects of supergravity and string theory on plane
wave backgrounds 
\be
ds^2 = 2 du dv + A_{ij}(u)x^i x^j du^2 + d\vec{x}^2\;\;.
\label{pwbc}
\ee
The main focus has been on time-independent plane waves, i.e.\ metrics
with $A_{ij}$ constant. For example, while any plane wave
solution of supergravity preserves half of the supersymmetries
\cite{chris}, these time-independent plane waves have been shown to
realise various exotic $> 1/2$ fractions of unbroken supersymmetries
\cite{pope,gh,michelson,benaroiban}. 

In another development \cite{mm} attention was drawn to {\em
homogeneous plane waves}. These generalise the time-independent (and
symmetric homogeneous) plane waves to time-dependent plane waves in a way
which does not destroy the homogeneity of the metric. These homogeneous
plane waves possess a number of interesting features and string theory
in these backgrounds has been studied in \cite{prt,mmga}.

Here we draw together these two, apparently unrelated, developments by showing 
\begin{itemize}
\item
that the existence of extra Killing spinors implies that the plane wave
metric is that of a smooth homogeneous plane wave (this corresponds to
one of the two families of homogeneous plane waves found in \cite{mm}),
\item
and (perhaps surprisingly) that such time-dependent plane waves can also
have extra Killing spinors and realise exotic fractions of supersymmetries.
\end{itemize}

The argument for the first claim is extremely simple and follows from
looking at the Killing vector constructed from the extra Killing spinor
(section 3.1).
To establish the second claim, in section 2 we study the Penrose \cite{penrose} 
limits of various G\"odel-like metrics and show that they give rise to non-trivial
homogeneous plane waves of the anti-Mach kind \cite{OS} studied in
\cite{mm,mmga}. In particular, the Penrose-G\"uven \cite{Gueven} limit
of the M-theory G\"odel metric discovered in \cite{chrisetal}, which
has 20 supersymmetries, is a time-dependent anti-Mach wave solution
of M-theory (section 2.3). Since the number of supersymmetries cannot decrease
in the Penrose limit \cite{bfp}, it must therefore be true that this
time-dependent plane wave admits extra Killing spinors, and we go on to
construct these Killing spinors explicitly 
(section 3.3).\footnote{This evidently contradicts
the ``proof'' in \cite{os,saka} that such solutions cannot exist. This proof 
rests on the invalid assumption that, as in the time-independent case 
\cite{gh}, also in the time-dependent case everything can be conjugated into the 
Cartan of $SO(16)$ in a time-independent way.} We also show that the 
existence of one extra Killing spinor implies the equations of motion (section 3.2), 
and we conclude with some general remarks on the Killing spinor
equations for homogeneous plane waves.

\section{G\"odel, Penrose, anti--Mach}

\subsection{A two-parameter family of G\"odel Metrics}

We first consider the two-parameter family
of G\"odel metrics (see e.g.\ \cite{bghv} and references therein)
\be
ds^2 = -(dt + \frac{4\sqrt{2}\Omega}{m^2} \sinh^2 \frac{m\rho}{2}d\phi)^2
+ d\rho^2 + \frac{1}{m^2}\sinh^2 m\rho\; d\phi^2 + dz^2
\label{gm1}
\ee
In terms of
\be
r = \sinh\frac{m\rho}{2}
\ee
and the parameters $\Omega$ and
\be
\Delta = \frac{4\Omega^2}{m^2}
\ee
this family can alternatively be written as
\be
ds^2 = - dt^2 - 2 \sqrt{2}\frac{\Delta}{\Omega} r^2 dtd\phi + 
\frac{\Delta}{\Omega^2} \frac{dr^2}{1+r^2}
+\frac{\Delta}{\Omega^2} (r^2 + (1-2\Delta)r^4) d\phi^2 + dz^2\;\;.
\label{gm2}
\ee
These metrics are homogeneous.
The orbits of the Killing vector $\del_{\phi}$ are closed, and since the norm
of $\del_{\phi}$ is proportional to $(1 + (1-2\Delta)r^2)$, one sees that there
are closed timelike curves for
\be
r^2 > \frac{1}{2\Delta -1}\;\;.
\ee
Thus these metrics share all the characteristics of the standard one-parameter
family of G\"odel metrics \cite{goedel} 
(with rotation parameter $\Omega$) which one obtains
for $\Delta = 1$. The one-parameter family $\Omega^2 = \Delta$ was shown in
\cite{roospin} to interpolate between the G\"odel metric at $\Delta=1$ and
the $AdS_3 \times \RR$ metric at $\Delta=1/2$ (where closed timelike curves 
are pushed to infinity and cease to exist). 

Another interesting limit is $m\ra 0$ with $\Omega$ fixed. 
Alternatively, in terms of the metric (\ref{gm2}) one scales 
\be
r^2 \ra \frac{\Delta}{\Omega^2} r^2
\ee
and then takes the limit $\Delta\ra\infty$. Then one  
finds the one-parameter family of metrics
\bea
ds^2 &=& - dt^2 - 2 \beta r^2 dt d\phi + dr^2 + (r^2 - \beta^2 r^4) d\phi^2 + dz^2\non
     &=& - (dt + \beta r^2 d\phi)^2 + dr^2 + r^2 d\phi^2 + dz^2
\label{gm3}
\eea
($\beta = \sqrt{2}\Omega$)
which is of the typical form of G\"odel-like metrics encountered in string
theory and M-theory \cite{chrisetal,thtt}. We will refer to it below as a
stringy G\"odel metric.

\subsection{The Penrose Limit of the stringy four-dimensional G\"odel Metric}

We now consider the Penrose limits of the above metrics. Before embarking
on the calculation let us try to anticipate what sort of result we expect.
As mentioned above, the G\"odel metrics are homogeneous. Now it was
pointed out in \cite{patricot} (by way of example) that the Penrose
limit of a homogeneous space-time need not be homogeneous. However,
it appears to be possible to show \cite{bfs} that the Penrose limit of
a homogeneous reductive space is itself homogeneous reductive. It is not
difficult to see that the G\"odel metrics are actually homogeneous reductive
(while the homogeneous Kaigorodov spaces considered in \cite{patricot}
are not). Thus we expect the Penrose limit of these G\"odel metrics to
be homogeneous plane waves.  This expectation will be borne out by our
explicit calculations below.

As it will turn out that the Penrose limits of all of the above metrics 
can also be obtained by starting from the simpler metric (\ref{gm3})
with $\beta=1$ (see Appendix A.2), we will just look at null geodesics 
in this case. 

The obvious Killing vectors $\del_z,\del_t,\del_\phi$ 
of (\ref{gm3}) give rise to the first integrals
\bea
\dot{z} &=& P \non
\dot{t} &=& 1- r^2\non
\dot{\phi} &=& 1
\label{ztf}
\eea
where an overdot denotes differentiation with respect to the affine parameter
$\tau$. Without loss of generality we have set the angular momentum $p_\phi=L$ to
zero. Then, using the remaining Killing vectors (Appendix A.1) it is not difficult 
to show (and in any case straightforward to verify) that the solution for $r(\tau)$ 
can be chosen to be
\be
r(\tau) = (1-P^2)^{1/2} \sin \tau\;\;.
\label{rt1}
\ee
To take the Penrose limit along any of the above geodesics (parametrised
by $P$), we change coordinates from $x^{\mu}=(r,z,\phi,t)$ to adapted
coordinates \cite{penrose,Gueven,bfp} $(u,v,y_1,y_2)$ with $u=\tau$. 
A possible choice is 
\bea
r &=& r(u)\non
dz &=& dy_1 + P du\non
d\phi &=& dy_2 + du\non
dt &=& - dv + P dy_1 + (1-r(u)^2) du\;\;.
\label{ac1}
\eea
Then, following the standard procedure, one finds the Penrose limit
\bea
ds^2 &=& 2 dudv + (1-P^2)dy_1^2 - 2P r(u)^2 dy_1 dy_2 + (r(u)^2 - r(u)^4)dy_2^2\non
r(u)^2 &=& (1-P^2) \sin^2 u\;\;.
\label{pl1}
\eea
The same one-parameter family of plane wave metrics is also obtained from the
Penrose limits of the general two-parameter family of G\"odel metrics (\ref{gm2}).

As usual, the metric in Rosen coordinates is not particularly revealing.
In particular, it is not obvious at this point that this is really a homogeneous 
plane wave. To exhibit this, we now show that we can put the above metric into 
the general form of a smooth homogeneous plane wave in stationary coordinates, namely
\cite{mm}
\be
ds^2 = 2 dudv + A_{ij}x^ix^j du^2 + 2 f_{ij}x^i dx^j du + d\vec{x}^2
\label{hpw}
\ee
with constant symmetric and anti-symmetric matrices $A_{ij}$ and $f_{ij}$
respectively.
 
We first rescale $y_1$ and $y_2$ by $(1-P^2)^{1/2}$ so that the metric
takes the form (reorganising the terms)
\be
ds^2 = 2 dudv + (dy_1 - P\sin^2 u dy_2)^2 + \frac{1}{4}\sin^2 2u dy_2^2\;\;.
\ee
We can deal with the second term in the standard way \cite{Gueven,bfp},
introducing 
\be
x_2 = \frac{1}{2} y_2 \sin 2u \;\;,
\label{x2}
\ee
and shifting $v$ appropriately to eliminate the $dudx_2$ cross-term. This will
have the net effect of generating $dx_2^2 -4 x_2^2 du^2$.
Instead of $y_1$ we introduce the coordinate
\be
x_1 = y_1 - P y_2 \sin^2 u \;\;,
\label{x1}
\ee
so that
\be
dy_1 - P\sin^2 u dy_2 = dx_1 + 2P x_2 du
\ee
Then one finds the metric
\be
ds^2 = 2 dudv + 4(P^2 -1) x_2^2 + 4P x_2 dx_1 du  + dx_1^2 + dx_2^2 
\;\;.
\ee
Finally, one more shift of $v$ to effect a ``gauge transformation'' of the 
magnetic field term puts the metric into the form
\be
ds^2 = 2 dudv + 4(P^2 -1) x_2^2 du^2 -2P (x_1dx_2- x_2 dx_1) du  + dx_1^2 + dx_2^2 
\;\;.
\ee
This shows that
the Penrose limit of the general G\"odel metric is indeed a homogeneous plane wave,
with 
\be
A_{11}=0\;\;,\;\;\;\;\;\;
A_{22}=4(P^2-1)\;\;,\;\;\;\;\;\;
f_{12} = -P\;\;. 
\ee
In particular, this one-parameter family of homogeneous plane waves is 
precisely of the anti-Mach kind \cite{OS} discussed in \cite{mm,mmga} 
in the sense that in stationary coordinates one of the frequencies is zero,
reflecting an additional commuting isometry. 

In summary: the Penrose limit of a G\"odel metric is an anti-Mach homogeneous
plane wave. 

\subsection{The Penrose Limit of the M-theory G\"odel Metric}

The five-dimensional G\"odel metric
\be
ds_5^2 = -(dt + \beta(r_1^2 d\phi_1^2 + r_2^2 d\phi_2^2))^2 
+ dr_1^2 + r_1^2 d\phi_1^2 + dr_2^2 + r_2^2 d\phi_2^2\;\;.
\label{g5}
\ee
has the remarkable property of being a maximally supersymmetric
solution of minimal five-dimensional supergravity (it preserves eight supercharges).
And it has the equally remarkable property that its M-theory lift
\be
ds_M^2 = ds_5^2 + dz^2 + \sum_{i=5}^{9} dx_i^2
\ee
(we have singled out one of the new six transverse dimensions)
supported by the four-form field strength
\be
F_{r_1\phi_156}= F_{r_1\phi_178}= F_{r_1\phi_19z}= F_{r_2\phi_256}= F_{r_2\phi_278}=
F_{r_2\phi_29z}=-2\beta\;\;,
\ee
preserves not only eight but actually 20 of the 32 supercharges of eleven-dimensional
supergravity \cite{chrisetal}. 
The analysis of null geodesics in this metric \cite{bghv}, using the
isometry algebra determined in \cite{chrisetal},  as well as the subsequent
Penrose limit of the metric, proceed in close analogy with the four-dimensional 
case discussed above, and we will be brief. 

We choose the null geodesic to have momentum $P$ along one of the six directions
transverse to the five-dimensional G\"odel metric which, without loss of generality,
we can choose to be the $z$-direction. Then similar to (\ref{ztf}) we have
\bea
\dot{z} &=& P \non
\dot{t} &=& 1- (r_1^2+r_2^2)\non
\dot{\phi_1} &=& \dot{\phi_2} = 1\;\;.
\eea
where
\be
r_1(\tau)^2 + r_2(\tau)^2 = (1-P^2) \sin^2 \tau
\label{rt2}
\ee
replaces (\ref{rt1}). This we solve as
\bea
r_1(\tau,\alpha)&=& (1-P^2)^{1/2} \cos \alpha \sin \tau \non
r_2(\tau,\alpha)&=& (1-P^2)^{1/2} \sin \alpha \sin \tau \;\;,
\label{ri}
\eea
where $\alpha$ is the angle between $r_1$ and
$r_2$ \cite{bghv}. The analogous adapted coordinates are (cf.\ (\ref{ac1}))
$(u,v,y_1,\tilde{\phi}_{1},\tilde{\phi}_2,\alpha)$ defined by
\bea
r_i &=& r_i(u,\alpha)\non
dz &=& dy_1 + P du\non
d\phi &=& d\tilde{\phi}_i + du\non
dt &=& - dv + P dy_1 + (1-(r_1(u)^2+r_2(u)^2)) du\;\;.
\label{ac2}
\eea
We take the Penrose limit along a geodesic sitting at
$\alpha=\alpha_0$ and hence introduce a new coordinate $y_3$ ($y_2$
will appear shortly \ldots) via
\be
\alpha = y_3 + \alpha_0\;\;.
\ee
Then one finds from (\ref{g5}) that the Penrose limit metric is
\bea
ds^2 &=& 2 dudv + (1-P^2)dy_1^2 + (r_{1}(u,\alpha_0)^2 + r_{2}(u,\alpha_0)^2)dy_3^2\non &&
+(r_1(u,\alpha_0)^2 - r_1(u,\alpha_0)^4)d\tilde{\phi}_{1}^2
+(r_2(u,\alpha_0)^2 - r_2(u,\alpha_0)^4)d\tilde{\phi}_{2}^2\non &&
- 2P r_1(u,\alpha_0)^2 dy_1 d\tilde{\phi}_{1} 
- 2P r_2(u,\alpha_0)^2 dy_1 d\tilde{\phi}_{2} \non &&
- 2 r_1(u,\alpha_0)^2 r_2(u,\alpha_0)^2 d\tilde{\phi}_1 d\tilde{\phi}_2 
+ \sum_{i=5}^9 dx_i^2
\;\;.
\eea
To disentangle this (and eliminate the apparent dependence of the metric on
$\alpha_0$) one can introduce the coordinates
\bea
y_2 &=& \cos^2 \alpha_0 \tilde{\phi}_{1} + \sin^2 \alpha_0 \tilde{\phi}_{2}\non
y_4 &=& \sin \alpha_0 \cos\alpha_0 (\tilde{\phi}_{2} -\tilde{\phi}_{1})\;\;.
\eea
Then the metric takes the form
\bea
ds^2 &=& 2 dudv + (1-P^2)\sin^2 u (dy_3^2 + dy_4^2)
+ \sum_{i=5}^9 dx_i^2 \non &&
 + (1-P^2)dy_1^2 - 2P r(u)^2 dy_1 dy_2 + (r(u)^2 - r(u)^4)dy_2^2
\eea
where, as in (\ref{pl1}), 
\be
r(u)^2 = (1-P^2) \sin^2 u\;\;.
\ee
We see that the resulting metric is quite simple: it has the form of a standard
symmetric (Cahen-Wallach) plane wave in the $(y_3,y_4)$-directions, and 
of the four-dimensional anti-Mach plane wave (\ref{pl1}) in the 
$(y_1,y_2)$-directions (times a flat $\RR^5$). 

We go to stationary coordinates as before. First we scale all
the $y_i$ by $(1-P^2)^{1/2}$. Then we go to Brinkmann coordinates
$x_{3,4}$ for the $y_{3,4}$-directions,
\be
x_{3,4}=\sin u y_{3,4}
\ee
and to coordinates $x_{1,2}$ as in (\ref{x1},\ref{x2}),
\bea
x_1 &=& y_1 - P y_2 \sin^2 u \non
x_2 &=& \frac{1}{2} y_2 \sin 2u \;\;.
\eea
(all this accompanied by an appropriate shift in $v$). 
Then one finds the metric
\be
ds^2 = 2 dudv + (4(P^2 -1) x_2^2 - x_3^2 -x_4^2) du^2 -2P (x_1dx_2- x_2 dx_1) du  + 
\sum_{i=1}^{9} dx_i^2\;\;.
\label{finally}
\ee
This is a non-trivial homogeneous plane wave with
\bea
A_{22} &=& 4(P^2 - 1)\non
A_{33} &=& A_{44} = -1\non
f_{12}&=& -P \;\;.
\label{amA}
\eea
As a check on this, note that for $P=0$ the null geodesic lies entirely
in the five-dimensional G\"odel metric, and must therefore lead to either 
the five-dimensional maximally supersymmetric plane wave \cite{patrick}
times $\RR^6$ (which also has 20 supersymmetries \cite{gh}) or flat space.
By inspection, one sees that it is the former.

It remains to determine the four-form field strength in the Penrose limit.
Using G\"uven's prescription \cite{Gueven,bfp} for taking the Penrose limit 
of supergravity fields other than the metric, and tracing through the chain
of coordinate transformations required to put the resulting metric into the
simple form (\ref{finally}), one finds that
\be
F_4 = 2 du\wedge [-dx^{129} + P dx^{349} - (1-P^2)^{1/2}(dx^{256} + dx^{278})]
\label{amF}
\;\;.
\ee
Here we used the shorthand notation
\be
dx^{abc} = dx^a \wedge dx^b \wedge dx^c\;\;.
\ee
As another check one can verify that the above metric and $F_4$ indeed satisfy the
supergravity equations of motion (\ref{eom}). 
Note that, even though the metric is trivial in
the $(x_5,\ldots,x_9)$-plane, there is non-trivial flux in those directions. 
In section 3.3 we will show by explicit construction that this supergravity
configuration has 20 (thus 4 extra) supersymmetries.
 
\section{Extra Killing Spinors and Homogeneous Plane Waves}

\subsection{Extra Killing Spinors $\Ra$ Homogeneity}

There is a very simple argument that shows that the existence of 
an extra Killing spinor implies that the plane wave is {\em homogeneous}.
A small refinement of this argument also shows that this homogeneous plane
wave must be smooth (corresponding to one of the two families of 
homogeneous plane waves found in \cite{mm}).

For definiteness we phrase the argument in the context of eleven-dimensional
supergravity, but it is clearly more general than that. We write the general
plane wave metric as 
\be
ds^2 = 2 dudv + A_{ij}(u)x^i x^j du^2 + d\vec{x}^2\;\;.
\ee
In a frame basis this metric is
\be
ds^2 = 2e^+e^- + (e^i)^2
\ee
where
\bea
e^+ &=& dv + \trac{1}{2}A_{ij}(u)x^i x^j du\non
e^- &=& du\non
e^i &=& dx^i\;\;.
\eea
It is well known \cite{chris} that any plane wave has 16 standard
supersymmetries, corresponding to Killing spinors $\epsilon$ with  
$\Gamma^-\epsilon=\Gamma_+ \epsilon =0$. Extra Killing spinors
are thus characterised by the condition
\be
\Gamma^- \epsilon \neq 0 \;\;.
\ee
Following the conventions of \cite{gh} we will also adopt in the following,
we choose
\be
\Gamma_{\pm} = \II_{16} \otimes \sigma_{\pm}
\label{gpm}
\ee
with $\sigma_{\pm} = (\sigma_1 \pm i\sigma_2)/\sqrt{2}$ so that
$(\Gamma_-)^T = \Gamma_+$.

Now consider the Killing vector\footnote{For a recent systematic discussion of
bispinors of eleven-dimensional supergravity see e.g.\ \cite{gapa}.} 
\be
K = \bar{\epsilon}\Gamma^M\epsilon \del_M\;\;.
\ee
Since $\Gamma^- = \Gamma^u$, it is clear that standard Killing spinors
can never give rise to a Killing vector with a non-zero $\del_u$-component,
in agreement with the fact that generic plane waves do not have such 
a Killing vector. The $\del_u$-component of $K$ for an extra
Killing spinor is
\be
K^u = \bar{\epsilon}\Gamma^-\epsilon = \epsilon^T C \Gamma^-\epsilon\;\;,
\ee
where $C$ is the charge conjugation matrix. $C$ can be chosen to be
$\Gamma^0$, where $0$ is a frame index, and thus 
\be
C = \trac{1}{\sqrt{2}}(\Gamma^+ - \Gamma^-)\;\;.
\ee
Then we have
\be
K^u = \trac{1}{\sqrt{2}}\epsilon^T \Gamma^+ \Gamma^-\epsilon
= \trac{1}{\sqrt{2}}(\Gamma^-\epsilon)^T (\Gamma^-\epsilon) \neq 0\;\;.
\ee
This shows that plane waves admitting extra Killing spinors have a 
Killing vector with a non-zero $\del_u$-component, i.e.\ they are homogeneous
\cite{mm}.

In \cite{mm} it was shown that there are two families of homogeneous plane
waves, smooth homogeneous plane waves which generalise the symmetric
(Cahen-Wallach, constant $A_{ij}$) plane waves but are generically
time-dependent, $A_{ij}=A_{ij}(u)$, and singular homogeneous plane waves
which generalise the metrics with $A_{ij}\sim u^{-2}$.
In Brinkmann coordinates the extra Killing vector $K$ has the form 
\be
K = \del_u + (\del_i-\mbox{pieces})
\ee
for smooth homogeneous plane waves, and the form
\be
K = u\del_u - v\del_v + (\del_i-\mbox{pieces})
\ee
in the singular case. Thus for singular plane waves the Killing vector
depends explicitly on $v$. Since it is easy to see that Killing spinors 
(be they standard or ``extra'') can never depend on $v$, this implies
that singular plane waves can have no extra supersymmetries. In 
particular, the existence of extra Killing spinors implies geodesic 
completeness.

In summary, the existence of an extra Killing spinor implies that the
plane wave is a smooth homogeneous plane wave. We will present the Killing
spinor equations for such metrics in the next section. 

\subsection{The Killing Spinor Equation for Homogeneous Plane Waves}

In \cite{mm} it was shown that the most general smooth homogeneous
plane wave can be written in stationary coordinates as (cf.\ (\ref{hpw})) 
\be
ds^2 = 2 dudv + A_{ij}x^i x^j du^2 + 2f_{ij}x^i dx^j du + dx^idx^i\;\;,
\ee
where $A_{ij}$ and $f_{ij}$ are constant symmetric and anti-symmetric matrices
respectively. In the standard Brinkmann coordinates (\ref{pwbc}) this
metric is explicitly time-dependent unless $A_{ij}$ and $f_{ij}$ commute.
We will specialise to the particular metric that arises in the 
Penrose limit of the M-theory G\"{o}del metric below. 

An orthonormal frame is
\bea
e^+ &=& dv + \frac{1}{2}A_{ij}x^ix^jdu + f_{ij} x^idx^k\non
e^- &=& du\non
e^i &=& dx^i\;\;.
\eea
The non-zero components of the spin connection are then
\be
\omega^{+i} = A_{ij} x^j du + f_{ij}dx^j\quad \omega^{ij} = -f_{ij} du
\ee
The Killing spinor equations for M-theory are
\be
(\nabla_M - \Omega_M)\epsilon = 0
\ee
where the covariant derivatives are those obtained from the above 
spin-connection, and the $\Omega_M$ are the contributions from the 
four-form field strength, 
\be
\Omega_M = \frac{1}{288}(\Gamma_M^{\;PQRS} - 8\delta^P_M\Gamma^{QRS})F_{PQRS}
\ee
We restrict $F$ to be of the homogeneous plane-wave form 
\be
F = \frac{1}{3!}du\wedge \xi_{ijk}dx^{ijk}
\ee
with constant $\xi_{ijk}$. Then the $\Omega_M$ are
\bea
\Omega_v &=& 0\non
\Omega_u &=& -\frac{1}{12} \Theta(\Gamma_+\Gamma_- + 1)\non
\Omega_k &=& \frac{1}{24}(3\Theta\Gamma_k + \Gamma_k\Theta)\Gamma_+
\eea
To economise notation we use the definitions 
\be
\Theta = \frac{1}{3!} \xi_{ijk}\Gamma^{ijk}\quad \quad 
\Phi = \frac{1}{2} f_{ij}\Gamma^{ij}\;\;.
\ee
Acting on spinors the covariant derivatives are
\bea
\nabla_v &=& \partial_v\non
\nabla_u &=& \partial_u - \frac{1}{2} \Phi - \frac{1}{2} A_{ij}x^j 
\Gamma_i\Gamma_+\non
\nabla_i &=& \partial_i - \frac{1}{2} f_{ji}\Gamma_j\Gamma_+
\eea
and the Killing spinor equations become
\bea
\partial_v\epsilon &=& 0\label{kveq}\\
\partial_u\epsilon &=& (\frac{1}{2}\Phi + \frac{1}{2} A_{ij}x^j 
\Gamma_i\Gamma_+ -\frac{1}{12}\Theta(\Gamma_+\Gamma_- + 1) )\epsilon \label{kueq}\\
\partial_i\epsilon &=& \tilde{\Omega}_i \epsilon \label{kieq}
\eea
where we have introduced 
\be
\tilde{\Omega}_i = \Omega_i + \frac{1}{2}\Gamma_if_{ik}\Gamma_+
\ee
We follow the analysis of \cite{fop,bfhp1}, as adapted to the non-maximally
supersymmetric case in \cite{pope,gh}.
The first equation (\ref{kveq}) implies that Killing spinors are independent of
$v$. Since $\tilde{\Omega_i}\tilde{\Omega_j}=0$, the third equation can immediately
be integrated to 
\be
\epsilon(u,x^i) = (1 + \sum x^i \tilde{\Omega_i}) \chi(u)
\ee
If $\epsilon(u,x^i)$ is such that it is annihilated by $\Gamma_+$, then it is
independent of the $x^i$, and from (\ref{kueq}) one finds the 16 standard Killing
spinors 
\be
\epsilon(u)=\ex{-\frac{1}{4}(\Theta - 2\Phi)u}\epsilon_0\;\;,\;\;\;\;\;\;
\Gamma_+\epsilon_0 =0
\ee
(where $\epsilon_0$ is a constant spinor)
which account for the $1/2$ supersymmetry of a generic plane wave. 

In the following we want to consider the possibility of extra supersymmetries
for which $\Gamma_+\epsilon\neq0$. 
Plugging $\epsilon(u,x^i)$ into (\ref{kueq}), we find a 
part that is independent of $x^i$ and
a part that is linear in $x^i$. We therefore solve separately the 
two parts of this equation. The $x^i$-independent part gives the equation
\be
\partial_u\chi(u) = 
(\frac{1}{2}\Phi - \frac{1}{12}\Theta(\Gamma_+\Gamma_- + 1))\chi(u)
\label{chiu}
\ee
which when substituted into the remainder, removing the overall factor of
$x^i$, gives the equations
\be
(\frac{1}{2}A_{ik} \Gamma_i\Gamma_+ + [\frac{1}{2}\Phi 
- \frac{1}{12}\Theta (\Gamma_+\Gamma_- + 1),\tilde{\Omega}_k])\chi = 0
\ee
Substituting the explicit expression for $\tilde{\Omega}_k$ into this equation
and multiplying by $288\Gamma_k$ (no summation over $k$) we find
\bea
&&(9\Gamma_k\Theta^2\Gamma_k + 6\Gamma_k\Theta\Gamma_k\Theta 
+ \Theta^2 
- 18\Gamma_k[\Phi,\Theta]\Gamma_k\non&& - 6[\Phi,\Theta] 
- 144 (A_{ik} +  f_{ij}f_{jk})\Gamma_k\Gamma_i)\Gamma_+ \chi(u) = 0\;\;.
\label{meq}
\eea
Note that the combination $A_{ik} +  f_{ij}f_{jk}$ that has popped up here is
essentially the Riemann tensor of the homogeneous plane wave.
We see that this is an equation for $\Gamma_+\chi(u)$ only. Solving (\ref{chiu}),
we find
\be
\Gamma_+\chi(u) = \Gamma_+\ex{\frac{1}{2}u(\Phi - 
\frac{1}{6}\Theta(\Gamma_+\Gamma_- + 1))}\chi_0 \non
= 
\ex{\frac{u}{2}(\Phi + \frac{1}{6}\Theta)}\Gamma_+\chi_0\;\;.
\ee
At this point it is convenient to switch to $SO(9)$ $\gamma$-matrices 
$\gamma_k$ via
\be
\Gamma_k = \gamma_k \otimes \sigma_3
\ee
with
\be
\gamma^{12\ldots 9}=\II_{16}
\ee
and with $\Gamma_\pm$ defined in (\ref{gpm}). Then we can equivalently 
write the condition (\ref{meq}) for the existence of extra Killing spinors as
\be
M_k\eta(u)=0
\label{meq2}
\ee
where
\be
M_k = 9\gamma_k\theta^2\gamma_k + 6\gamma_k\theta\gamma_k\theta 
+ \theta^2 
- 18\gamma_k[\phi,\theta]\gamma_k - 6[\phi,\theta] 
- 144 (A_{ik} +  f_{ij}f_{jk})\gamma_k\gamma_i\;\;,
\label{mk}
\ee
with
\be
\theta = \frac{1}{3!} \xi_{ijk}\gamma^{ijk}\quad,\quad
\phi = \frac{1}{2} f_{ij}\gamma^{ij}\;\;,
\ee
and
\be
\eta(u) = \ex{\frac{u}{2}(\phi + \frac{1}{6}\theta)}\eta_0
\label{etau}
\ee
a 16-component spinor.

Simple algebraic manipulations show that 
$\sum_k M_k$ is proportional to the identity matrix times the
lhs of the equation of motion 
\be
\tr (A + f^2) + \trac{1}{12} \xi_{ijk}\xi^{ijk}=0\;\;.
\label{eom}
\ee
Thus the existence of just one extra Killing spinor is sufficient to
guarantee that the equations of motion are satisfied.

\subsection{The Extra Killing Spinors of the M-theory anti--Mach Plane Wave}

Since the number of supersymmetries can never decrease in the Penrose
limit \cite{bfp}, we expect the M-theory anti-Mach metric (\ref{finally})
which we obtained as the Penrose limit of the M-theory G\"odel metric
to possess (at least four) extra supersymmetries. We will now verify
this explicitly.

For the M-theory anti-Mach metric, the non-zero components of $\xi$ are
(\ref{amF})
\bea
\xi_{129} &=& -2\non
\xi_{349} &=& 2P\non
\xi_{256} &=& -2(1-P^2)^{\frac{1}{2}}\non
\xi_{278} &=& -2(1-P^2)^{\frac{1}{2}}\non
\label{xixi}
\eea
and the non-zero components of $A_{ij}$ and $f_{ij}$ are (\ref{amA})
\bea
A_{22} &=& 4(P^2 - 1)\non
A_{33} &=& A_{44} = -1\non
f_{12}&=& -P
\eea

Taking into account the obvious symmetries of the metric and field strength it 
is sufficient to consider in detail only the $k=1,2,3,5,9$ components 
of (\ref{meq2}).
For convenience we will write $P = s = \sin\theta$ 
and hence also $(1-P^2)^{\frac{1}{2}}= c = \cos\theta$. 
The $M_k$ of interest are (we do not write the identity matrix explicitly in the
following)
\begin{itemize}
\item
k=1
\be
- 3 + 5s^2 + 2s\gamma_{1234} + 2c\gamma_{1956} + 2c\gamma_{1978} + (1-s^2)
\gamma_{5678} - 3sc(\gamma_{156} + \gamma_{178})
\ee
\item
k=2
\be
12 - 10s^2 + 2s\gamma_{1234} - 4c\gamma_{1956} - 4c\gamma_{1978} 
+ 4(1-s^2)\gamma_{5678} + 3sc(\gamma_{156} + \gamma_{178})
\ee
\item
k=3
\be
3 - s^2 + 2s\gamma_{1234} - c \gamma_{1956} - c \gamma_{1978} 
+ (1-s^2) \gamma_{5678}
\ee
\item
k=5
\be
-6 + 4 s^2 - 2 s \gamma_{1234} + 4 c \gamma_{1956} - 2 c \gamma_{1978} 
- 4 (1 - s^2) \gamma_{5678} + 3sc(\gamma_{156} - \gamma_{178})
\ee
\item
k=9
\be
3 + s^2 + 4s\gamma_{1234} - 2c \gamma_{1956} - 2c\gamma_{1978} 
- (1-s^2) \gamma_{5678}
\ee
\end{itemize}

With a little algebra one can easily show that all of these expressions
have the general form
\be
M_k = A_k (1 + \gamma_{5678}) + B_k (1 - c\gamma_{1956} + s\gamma_{1234})
+ C_k (1 - c \gamma_{1978} + s\gamma_{1234})\;\;,
\ee
where the coefficients $A_k,B_k,C_k$ can include other gamma matrices
that are not important for our discussion. It is then easily checked that 
\be
M_k P_1 P_2 = 0
\ee
for all $k$, where
\bea
P_1 &=& \frac{1}{2}(1 - \gamma_{5678}) \non
P_2 &=& \frac{1}{2} (1 + c \gamma_{1956} - s \gamma_{1234})
\eea
are two commuting projection operators, 
\be
P_1^2 = P_1\;\;,\;\;\;\;P_2^2 =P_2\;\;,\;\;\;\;P_1P_2 = P_2 P_1\;\;. 
\ee
Moreover, one can check that $P_1P_2$ commutes (!) with 
$\phi + \frac{1}{6}\theta$, so that the extra Killing spinors
are (cf.\ (\ref{etau}))
\be
\eta(u) = \ex{\frac{u}{2}(\phi + \frac{1}{6}\theta)}P_1P_2\eta_0\;\;.
\ee
This gives precisely four extra Killing spinors, consistent with the
20 supersymmetries of the M-theory G\"odel metric. We have thus demonstrated
explicitly that time-dependent (homogeneous) plane waves can admit extra
supersymmetries.

For $s=0$, i.e.\ $P=0$, one reproduces the result 
of \cite{gh} that the M-theory lift of the maximally supersymmetric
five-dimensional plane wave \cite{patrick} has 20 supersymmetries.
In this case the projectors 
\bea
P_1 &=& \frac{1}{2}(1 - \gamma_{5678}) \non
P_2 &=& \frac{1}{2} (1 + \gamma_{1956})
\eea
are constructed in the standard way from commuting four-vectors
$\gamma^{(4)}$ of the Clifford algbra. 

Something interesting happens when we switch on $s=P$ or $f_{ij}$,
i.e.\ when we switch on a time-dependence of the homogeneous plane
wave (in Brinkmann coordinates).  It is clear from (\ref{xixi}) that
this has the effect of switching on a component in $\theta$, namely
$2P\gamma_{349}$, which does not commute with all the other components.
As a consequence the corresponding projectors cannot be built anymore
from commuting elements $\gamma^{(4)}$ of the Clifford algebra alone -
and indeed for $P_2$ to be a projector for $s\neq 0$ it is essential
that $\gamma_{1956}$ and $\gamma_{1234}$ anti-commute!

This is clearly a general feature of extra Killing spinors of non-trivial
($f_{ij}\neq 0$) homogeneous plane waves that gives the analysis a rather
different flavour. In particular, while for time-independent plane waves
one can diagonalise everything in sight, this is not the case in general.
This evidently complicates the analysis of the general solutions of the
equations (\ref{meq2}), and we will leave a more detailed investigation
of these and related issues for the future.

\subsection*{Acknowledgements}

MB is grateful to the SISSA High Energy Physics Sector for financial
support. The work of PM and MO is supported in part by the European
Community's Human Potential Programme under contracts HPRN-CT-2000-00131
and HPRN-CT-2000-00148.

\appendix

\section{More on the two-parameter family of G\"odel metrics}

\subsection{Null geodesics}

In addition to the obvious Killing vectors
$\del_{\phi},\del_t, \del_z$ the metric (\ref{gm2}) has the
two Killing vectors $K_i$, $i=1,2$
\be
K_i = (1+r^2)^{1/2} g_i(\phi)\del_r + \sqrt{2} \frac{\Delta}{\Omega}
\frac{r}{(1+r^2)^{1/2}} g_i^{\prime}(\phi)\del_t + \frac{1+2r^2}{r(1+r^2)^{1/2}}
g_i^{\prime}(\phi)\del_\phi\;\;,
\ee
where the $g_i$ are any two linearly independent solutions of the equation
\be
g^{\prime\prime}(\phi) + g(\phi) =0\;\;,
\ee
e.g.\ $g_1=\sin\phi$, $g_2 = \cos\phi$. There are obviously (more than)
enough isometries to determine the null geodesics completely. The Killing
vectors $\del_y,\del_\phi$ and $\del_z$ give us the first integrals (an
overdot denotes a derivative with respect to the affine parameter $\tau$)
\bea
E &=& \dot{t} + \sqrt{2}\frac{\Delta}{\Omega} r^2 \dot{\phi}\non
L &=& \frac{\Delta}{\Omega^2} (r^2 + (1-2\Delta)r^4) \dot{\phi}
-\sqrt{2}\frac{\Delta}{\Omega} r^2 \dot{t}\non
P &=& \dot{z}\;\;.
\eea
By a scaling of $\tau$ we can choose $E=1$.  Moreover, because of the
covariance of the Penrose limit \cite{bfp}, without loss of generality we
can choose $L=0$ (there are enough isometries to generate geodesics with
$L\neq 0$ from those with $L=0$).  Then the first two equations lead to
\bea
\dot{t} &=& \frac{1+(1-2\Delta)r^2}{1+r^2}\non
\dot{\phi} &=& \frac{\sqrt{2}\Omega}{1+r^2}\;\;.
\eea
To determine $r(\tau)$ we make use of the conserved charges $F_i$
associated with the Killing vectors $K_i$, which (with $L=0$, $E=1$) are
\be
F_i = 
\frac{\Delta}{\Omega^2}\frac{1}{(1+r^2)^{1/2}} g_i(\phi)\dot{r} 
- \sqrt{2} \frac{\Delta}{\Omega}
\frac{r}{(1+r^2)^{1/2}} g_i^{\prime}(\phi) \;\;.
\ee
Solving these for $\dot{r}$ and equating the resulting expressions,
one finds
\be
\sqrt{2} \frac{\Delta}{\Omega} \frac{r}{(1+r^2)^{1/2}} = F_2 \sin\phi - F_1 \cos\phi
\;\;.
\ee
Differentiating both sides and using the expression for $\dot{\phi}$ one obtains
\be
\frac{\Delta}{\Omega^2} \frac{\dot{r}}{(1+r^2)^{1/2}} = F_1 \sin\phi + F_2 \cos\phi
\;\;.
\ee
Hence squaring and adding the two equations one gets
\be
\frac{\Delta}{\Omega^2}\frac{\dot{r}^2}{1+r^2}
= \frac{\Omega^2}{\Delta}(F_1^2 + F_2^2) - 2\Delta \frac{r^2}{1+r^2}
\;\;.
\ee
Comparing this with the null constraint
\be
\frac{\Delta}{\Omega^2}\frac{\dot{r}^2}{1+r^2}
= \dot{t}^2 + 2 \sqrt{2}\frac{\Delta}{\Omega} r^2 \dot{t}\dot{\phi}  
-\frac{\Delta}{\Omega^2} (r^2 + (1-2\Delta)r^4) \dot{\phi}^2 - \dot{z}^2\;\;,
\ee
one finds that these two expresions are equal (for all $r$) provided that
the single constraint
\be
\frac{\Omega^2}{\Delta}(F_1^2 + F_2^2) = 1 - P^2
\ee
is satisfied. Using this to eliminate the $F_i$ in favour of $P$, 
we then find the solution
\be
r(\tau) = \left(\frac{1-P^2}{2\Delta + P^2 -1}\right)^{1/2}\sin \omega \tau\;\;,
\ee
where
\be
\omega = (2 +(P^2-1)/\Delta)^{1/2}\Omega\;\;.
\label{omega}
\ee
In particular, for the stringy G\"odel metric (\ref{gm3}) one finds
(either directly or by taking the limit)
\bea
\dot{t} &=& 1-\beta^2 r^2\non
\dot{\phi} &=& \beta\non
r(\tau) &=& (1-P^2)^{1/2} \frac{\sin \beta \tau}{\beta}\;\;.
\eea

\subsection{The Penrose Limit of four-dimensional G\"odel-like Metrics}

To take the Penrose limit of these G\"odel metrics along any of the above
geodesics, we go to adapted coordinates \cite{penrose,Gueven,bfp}, i.e.\ 
we seek a change of coordinates from $x^{\mu}=(r,z,\phi,t)$ to $(u,v,y_1,y_2)$ 
with $u=\tau$ in such a way that $g_{uv}=1$ and $g_{uu}=g_{u1}=g_{u2}=0$. 
We can e.g.\ choose
\bea
r &=& r(u)\non
dz &=& dy_1 + \dot{z}(u) du\non
d\phi &=& dy_2 + \dot{\phi}(u) du\non
dt &=& - dv + P dy_1 + \dot{t}(u) du\;\;.
\eea
The coordinate $v$ (the only one that may require some explanation)
can be found by either of the two methods employed in \cite{bfp}
or, more elegantly, using the Hamilton-Jacobi method advocated in \cite{patricot}.
In any case, one can check that this really is an adapted coordinate system,
and one can thus take the Penrose limit to find the plane wave metric
\be
ds^2 = 2 dudv + (1-P^2)dy_1^2 -2\sqrt{2}P \frac{\Delta}{\Omega}r(u)^2 dy_1dy_2
+ \frac{\Delta}{\Omega^2} (r(u)^2 + (1-2\Delta)r(u)^4)dy_2^2\;\;.
\ee
At this point, the metric appears to depend on the three parameters
$(\Delta,\Omega,P)$. It is not difficult, but a bit tedious, to see
that in fact all the dependence on $\Omega$ and $\Delta$ can be eliminated
by various scalings of the coordinates (and a redefinition of $P$).
First of all, by a scaling of $u$ and a reciprocal scaling of $v$ one can
eliminate $\omega$. Then the remaining dependence on $\Omega$ can be eliminated by a 
scaling of $y_2$. To deal with the $\Delta$-dependence, one writes 
\be
r(u)^2 + (1-2\Delta)r(u)^4 = (2\Delta-1)^{-1}(\tilde{r}(u)^2 - \tilde{r}(u)^4)
\;\;.
\ee
Further scalings of both coordinates can then be used to put the metric into the form
\bea
ds^2 &=& 2 dudv + (1-Q^2)dy_1^2 - 2Q r(u)^2 dy_1 dy_2 + (r(u)^2 - r(u)^4)dy_2^2\non
r(u)^2 &=& (1-Q^2) \sin^2 u\;\;,
\eea
where $Q$ is related to $P$ by
\be
Q = \sqrt{2}P \frac{\Delta^{1/2}}{(2\Delta + P^2 -1)^{1/2}}\;\;.
\ee
Note that $Q=0$ ($Q=1$) iff $P=0$ ($P=1$). 
This is precisely the one-parameter family of plane wave metrics one also obtains
from the simpler stringy G\"odel metric (\ref{gm3}) with $\beta =1$, and
$Q\ra P$. At this point the analysis proceeds as for this special case 
discussed in section 2.2.

\rnc{\Large}{\normalsize}

\end{document}